# MarioChart: Autonomous Tangibles as Active Proxy Interfaces for Embodied Casual Data Exploration

Shaozhang Dai
Monash University
Melbourne, Australia

Kadek Ananta Satriadi
Monash University
Melbourne, Australia

Jim Smiley
Monash University
Melbourne, Australia

Barrett Ens
The University of British Columbia
British Columbia, Canada

Lonni Besançon
Linköping University
Norrköping, Sweden

Tim Dwyer
Monash University
Melbourne, Australia

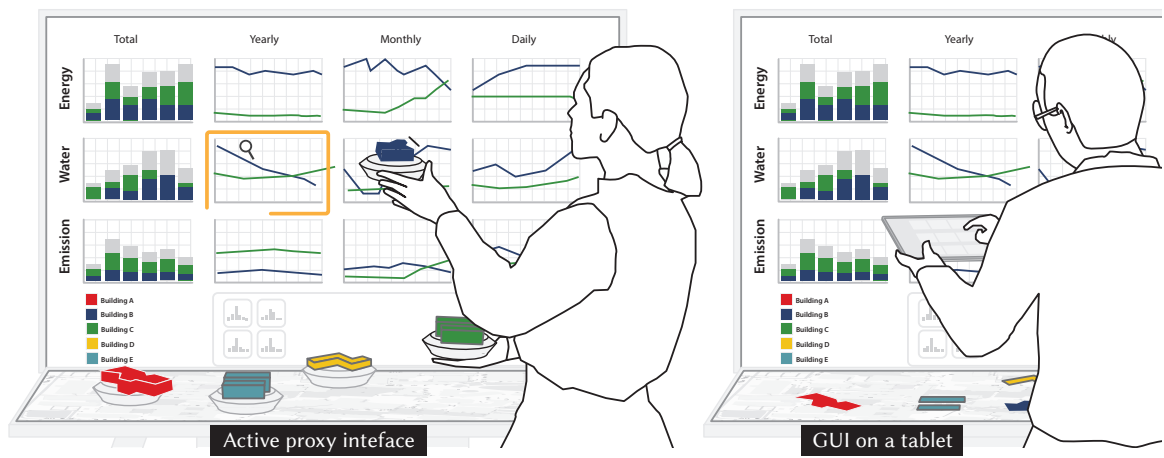

Figure 1: We investigate the strengths and weaknesses of tangible interactions using scale models of physical referents for visual data exploration, a paradigm we refer to as an Active Proxy interface. In the left figure, the user has picked up two building models to filter the charts on a dashboard and is moving one of them toward a chart of interest to drill down into more granular data. More such active-proxy interactions are described in the paper. We compare active-proxies against graphical user interfaces on a tablet.

## Abstract

We introduce the notion of an Active Proxy interface, i.e. tangible models as proxies for physical data referents, supporting interactive exploration of data through active manipulation. We realise an active proxy data visualisation system, "MarioChart", using robot carts relocating themselves on a tabletop, e.g., to align with their data referents in a map or other visual layout. We consider a casual-data exploration scenario involving a multivariate campus sustainability dataset, using scale models as proxies for their physical building data referents. Our empirical study (n=12) compares active proxy use with conventional tablet interaction, finding that our active proxy system enhances short-term spatial memory of data and enables faster completion of certain data analytic tasks.
It shows no significant differences compared to traditional touch-screens in long-term memory, physical fatigue, mental workload, or user engagement. Our study offers an initial baseline for active proxy techniques and advances understanding of tangible interfaces in situated data visualisation.

## ACM Reference Format:

# 1 Introduction

According to Ishii (2007), Tangible User Interfaces (TUIs) "give physical forms to digital information [which] serve as both representations and controls for their digital counterparts." [34] Recently, 3D printing and object tracking technology have made many TUI use cases feasible and relatively cheap. Tangibles can come in various forms and roles in a wide range of applications. They can function as user interface elements (e.g., menus [16], data flow diagrams [95], lenses [49], and more) or representations of information, such as data physicalization [5, 40] or scale models of physical objects. Inspired by such tangible interfaces, we explore how these concepts can be applied to situated visualisation contexts. Particularly, our research builds on situated visualisations that combine tangible scale models of physical objects with digital data representations [23, 68].

Situated visualisation and analytics is a data visualisation approach that incorporates the physical objects or spaces to which the data refers in the analysis process [20, 21, 81, 91, 92]. These objects or spaces are known as *physical referents* [92]. In situations where the referents are inaccessible, they can be represented using proxies, in scenarios known as *proxsituated* visualisation [67]. For example, the Uplift system [23] introduced tangible scale models (tangible proxies) for buildings arranged on a tabletop campus map (referent), which could be picked up to explore an energy usage dataset. They argued that exploring the dataset through interaction with these tangible proxies afforded a kind of "casual" data exploration that was approachable for non-expert users. Scenarios included information sharing between diverse stakeholders in meetings and engaging the public in energy use discussions through installation in building foyers. A key limitation of the Uplift system was that it required users to return building models to their correct positions on the map after interaction, or if the map is zoomed or panned.

Tangible interaction has been demonstrated to be beneficial in facilitating learning [69]. However, despite compelling system demonstrations such as Uplift, to our knowledge, no studies are exploring the effect of active manipulation of tangible proxies on visualisation task performance. This leaves a gap in our understanding of the benefits and drawbacks of tangibility in such scenarios. We ask two main questions: (1) How can an active proxy system be designed to be intuitive, taking into account that users are accustomed to traditional input devices and techniques? (2) Do the benefits of tangible interaction translate to interactive situated visualisations that rely on active manipulation of referent proxies such as scale models?

In addressing these questions, our paper presents three main contributions:

- We propose conceptual quadrants that can position existing and future work, as well as our own, as the product of two axes: (1) a passive *vs.* active proxy axis and (2) a spatial *vs.* abstract data representation axis. Our work focuses on active proxies for understanding data using abstract representations, such as a charts dashboard[1]. We elaborate further on this in section 3.

- A design and implementation of the active proxy system we called *MarioCharts* (section 4). Our MarioCharts system demonstrates the active proxy interface concept by allowing users to manipulate tangible scale models, position-tracked in six degrees of freedom, to explore data and visualisation associated with them. We also made the tangibles self-driving to facilitate automatic placements on the map after they are manipulated by the user or after the map is changed by pan or zoom. We explored design choices and realised a usable system to inform future work.

- An empirical evaluation of the active proxy interface (section 5). With 12 participants, we ran a user study comparing MarioChart with a graphical user interface (GUI) on a tablet. Our findings suggest that the active proxy interface enhances short-term spatial memory of data and enables faster completion of certain data analytic tasks than a GUI on a tablet. The active proxy interface has no significant differences compared to the tablet interface in terms of long-term memory, physical fatigue, mental load, and user engagement. Our study offers an initial baseline for the active proxy concept and advances understanding of tangible interfaces in situated data visualisation.

We found our findings, particularly regarding short-term memory improvements, are compelling evidence for the potential of TUIs for enabling casual data exploration following the active proxy concept. In section 9, we outline further use-cases for active-proxy data exploration in GLAM[2] and education contexts to motivate further study (section 10).

# 2 Related Work

Our work relates to two established topics in HCI: tangible user interfaces and situated visualisation. Specifically, we explore tangible interface concepts and interactions within situated visualisation scenarios that employ tangible proxies to represent the physical referents of data.

## 2.1 Tangible User Interfaces

Interaction via tangible user interfaces (TUIs) is a long-discussed topic in HCI. One of the seminal works is a monograph by Shaer et al. [71] published in 2010—two decades after TUI first emerged—that summarises TUI from multiple perspectives, such as tangible AR, tangible tabletop interaction, ambient display, and embodied user interfaces. TUI emerged from preceding notions of "graspable interfaces" [24] in 1995 and "tangible bits" by Ishii and Ullmer [36] in 1997. These concepts put forward an idea of physical world interfaces in which objects and surfaces connect users with digital data. As suggested by Shaer et al. [71], tangible interfaces and interaction can be understood from varying perspectives. Tangible interfaces adopt a data-centric paradigm emphasising device-based manipulation, whereas tangible interactions embrace an embodiment-centred approach that privileges physical skills and kinaesthetic expression. For this paper, we use the term TUI to encompass both tangible interface (the physical proxy and its data) and tangible interaction (how users manipulate the proxy), since our active proxy interface concept integrates both the device itself and its physical manipulation for chart interaction. Foundational TUI research encompasses early system implementations [84, 86], systematic literature analyses [10, 57, 66], theoretical frameworks [42, 45, 51, 72, 85], spatial interaction models [73], and canonical works [34]. As our research addresses TUI deployment in situated data visualisation scenarios, we proceed to examine empirical investigations of TUI efficacy and limitations.

---

[1]Sections of this paper defining our Active Proxy concept are derived from our CHI'23 late-breaking work presentation - see the anonymised version in supplementary material

[2]GLAM = Galleries, Libraries, Archives and Museums

Tangible-graphical interface comparisons trace back to Fitzmaurice and Buxton [25]'s graspable interface study, which showed tangible interfaces outperforming graphical ones in performance and usability. Empirical evaluations of TUIs emerged after that. The overall trend of tangible user interfaces (TUI) in terms of performance demonstrates superiority over graphical user interface (GUI) baselines across touch and mouse modalities [26, 37, 80, 83, 93]. Beyond performance metrics, TUIs exhibit cognitive advantages, including enhanced information recall [60], improved spatial cognition [43, 60, 65], and facilitated problem-solving and learning processes [43, 50]. Furthermore, existing research indicates user preference for TUI over GUI implementations [37, 83, 95], with additional evidence suggesting that TUI promotes collaborative learning environments [28].

It is crucial to emphasise that the contexts of these evaluations differ substantially from our work. Educational and learning applications represent the predominant theme, encompassing children's learning tools for floor plan design [43] and numerical learning [50], puzzle-based interactions [93], shape manipulation [83], physics education [37, 52], medical training [65], and visual programming education [28]. Another prominent theme involves creative and design tools, including drawing applications [26], document organisation systems [60], and photo management [80]. Several studies specifically focused on children's interactions [16, 50, 53]. Our research extends TUI applications to data visualisation contexts, exploring physical manipulation of tangible proxies for understanding associated data whilst revealing the dual roles of tangibles: 1) as representations of physical referents, and 2) as mechanisms for tangible interaction in data exploration. Our work investigates methods for communicating relationships between physical referents and data representations within casual data exploration scenarios. To our knowledge, empirical evaluation of TUIs within this specific context remains an underexplored area, leaving a gap in the strengths and weaknesses of TUIs for casual data exploration scenarios. We discuss it further in the following section.

## 2.2 Situated Visualisation with Tangibles

Situated visualisation and analytics situate data representations within the context of their physical referents and have been investigated for almost two decades [13, 14, 20, 20, 21, 47, 54, 74, 81, 90–92]. As suggested by existing work [67], there are situated visualisation scenarios in which immediate physical referents (e.g., buildings) are substituted by their proxies, such as tangible scale models, keeping a certain level of situatedness to the data visualisation and analysis process. This is the focus of our work, as it opens up opportunities for incorporating TUIs. Part of our interest includes understanding if physical manipulations of scale models to explore data form a strong conceptual binding between the referent and the data in the user's mind. The term "binding" is inspired by theories of learning [64], where binding is considered as a process of associating features of different entities in one's memory. Beyond information recall, common measures such as performance, accuracy, engagement, and preference are also of interest in our study.

While tangibles have been used in situated data visualisation, few studies have incorporated tangible proxies of physical referents. Most research employs tangibles combined with tabletop displays to represent interaction operators—such as opening views, changing visualisation idioms, filtering, and selection—rather than representing physical objects [1, 27, 30]. Related work includes interactive scatter plots [87], small tangible displays [17, 63], and visualisation design [32, 46]. Examples also exist in immersive analytics contexts [22, 76, 78, 82].

Situated visualisation with tangible proxies varies in the roles of tangibles within the visualisation process, the types of data representations visualised, application contexts, and other factors. The following section proposes a design space to describe these variations and discusses existing work within this domain using the proposed framework.

## 3 Design Space

We first propose a design space positioning situated visualisation work, including our own, that represents physical referents with tangible proxies such as physical scale models. There are two main factors (or dimensions) in our design space that differentiate the roles of proxies in interaction and data representation, as follows.

### 3.1 Passive vs. Active Proxy

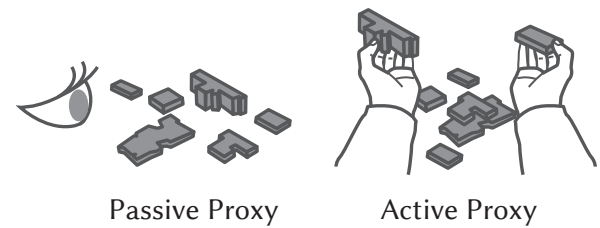

Figure 2: Tangible proxies could be visually perceived without physical manipulation (left) or actively manipulated (right).

The first factor describes the degree of the user's active engagement with proxies within the visualisation process, Figure 2. Passive proxies are characterised by predominantly static scale models; a typical example would be tangible maps with fixed-in-place building models. This approach emphasises the precise spatial arrangement of tangible proxies corresponding to their actual referent objects. In passive proxy scenarios, data manipulation is often performed through separate input devices. For instance, when selecting data attributes or modifying colour mapping, users interact with graphical user interface elements displayed on a screen. Active proxies are characterised by physical scale models that can be, or are specifically designed to be, actively manipulated. In active proxy scenarios, reliance on external input devices is minimised. Manipulation of data representation is performed exclusively through physical manipulation of the scale models. Naturally, the two scenarios are not mutually exclusive, as active proxy systems may also be used passively.

## 3.2 Spatial vs. Abstract Data Representation

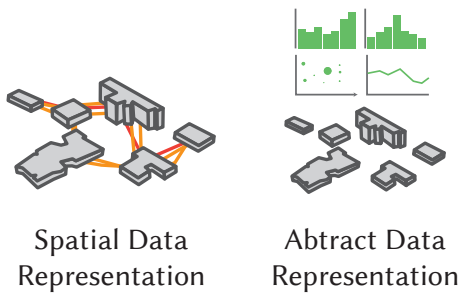

Figure 3: Spatial data representations pose more constraints (left) than abstract data representations (right).

The second factor describes the type of data representation employed (Figure 3). Spatial representation overlays data in its correct spatial context in the space around the tangible proxies. Complex spatial data representation is commonly presented as multiple geographical layers. Whilst this approach is highly spatially coupled with the referent, multiple layers can increase visual clutter. Abstract representation presents data that does not naturally integrate with the spatial context. For instance, time-series bar charts or line charts representing the energy consumption of campus buildings. Abstract data representations are spatially decoupled from the referent. Consequently, increasing the amount of presented information does not introduce the same occlusion issues encountered in spatial representation.

## 3.3 Quadrants

The interaction between the two factors above creates four quadrants (Figure 4) that can be used to position work in this domain. Table 1 shows how some of the existing work fits into the design space.

*3.3.1* **Passive Proxy** *with* **Spatial Representation**. This scenario uses static tangible models combined with spatial data overlaid on the surface. Examples include public installations of city visualisations (e.g., City of London[3], Nantes[4], Kuala Lumpur[5]). Another example is a tabletop system in which geovisualisation layers, such as road networks, are overlaid onto a physical terrain proxy [44]. Input devices – typically tablets or, in some instances, tangible interfaces – are employed to control the interactive features.

---

[3]https://nla.london/videos/new-london-model
[4]http://devocite.com/?page_id=1202
[5]https://cubexis.com.my/portfolio/scale-model-projection-mapping/

*3.3.2* **Passive Proxy** *with* **Abstract Representation**. This scenario uses static tangible models combined with abstract charts presented around the models. Examples include augmented reality works by Benkhelifa et al. [7] and Satriadi et al. [68] that explored effective placements of abstract data representation, such as bar charts, Gantt charts, scatterplots, and others, around passive tangible proxies placed on a table. This scenario can also be applied on interactive table top (see work by Walker et al. [88]).

*3.3.3* **Active Proxy** *with* **Spatial Representation**. This scenario uses tangible models to actively manipulate spatial data representation. Two common approaches for creating active proxies involve physical manipulations and physical deformations tangible proxies. Physical manipulation examples include the Urp system [86] that enables the physical arrangement of building models to create air flow visualisations. Similarly, Hull et al. [31] demonstrates a data exploration technique through the manipulation of scale models, where spatial information is presented as data layers. Physical deformation examples include deformation of tangible terrain to update geographic data visualisations (e.g. simulated water flow) based on the modified shape of 3D models deformed by users [35, 55, 61].

*3.3.4* **Active Proxy** *with* **Abstract Representation**. This scenario uses tangible models to actively manipulate abstract data representation. An exemplary work that fits well within this category is that of Jofre et al. [41]. Their system employs a dashboard displaying abstract data related to radio stations and listeners. The tangible proxies can be arranged on the table to query the data, thereby constructing different types of visualisation. Other examples within this domain exhibit certain levels of overlap with other quadrants. [2] and [94] utilise active proxies and display both spatial and abstract representations. The Uplift system by Ens et al. [23] forms part of this quadrant, but also incorporates static proxies and spatial data representation.

## 4 MarioChart System

The proposed design space describes various possible scenarios that reflect the interplay between how actively tangible proxies are manipulated by users and how to design the data representations. Motivated by the possibility for novel techniques offered by the active proxy combined with abstract data representation, we investigate further by designing and evaluating a system that we call MarioChar.t [6]. We designed MarioChart as an exploratory prototype to investigate how an active proxy system can support interaction with complex data visualisations. Instead of exploring raw dataset, the system aims to facilitate convenient navigation of multi-dimensional data within a situated visualisation environment, supporting tasks such as performing casual exploration, presenting information, or sharing insights. The interface is intentionally designed to be easy to learn, allowing users to interact with the system with minimal guidance. This section describes the workflow and interaction designs, system overview, and implementation details. Jansen and Dragicevic [38] propose an interaction model for data visualisation beyond the desktop interfaces, including tangible input. They use diagrammatic representations of the elements of their model to map the role of physical manipulations of tangible controls in various data visualisation and physicalisation use cases. Inspired by this work, Figure 5 maps the role of physical manipulation in the MarioChart system to data transformation, perception and information understanding in the MarioChart system.

---

[6]Inspired by the self-driving proxies that look like karts in Mario Kart games.

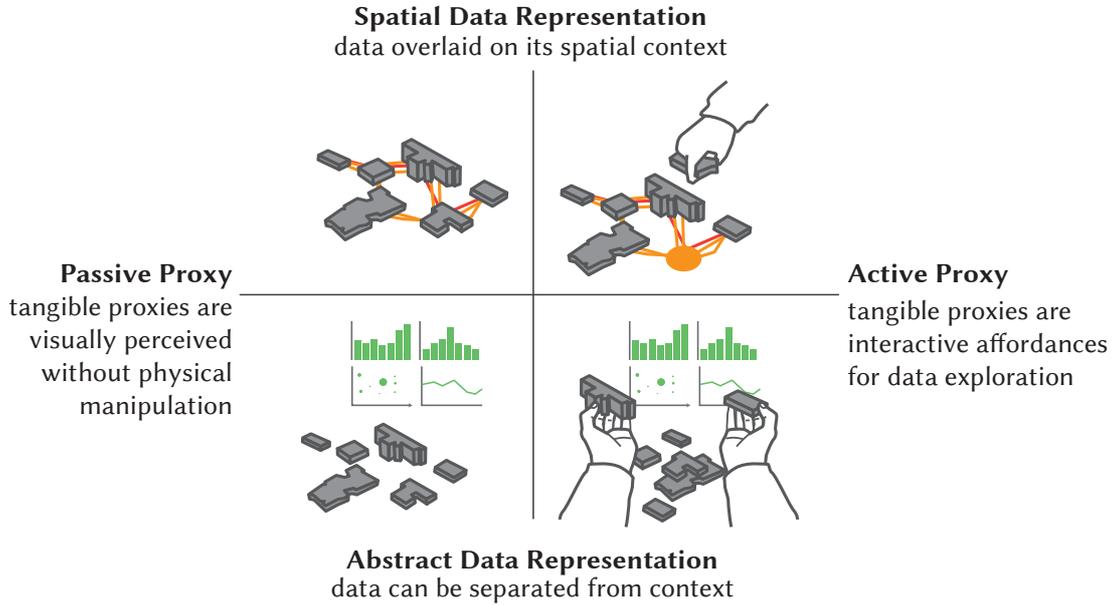

Figure 4: Conceptual quadrants showing passive versus active proxy and spatial versus abstract representation.

| Title | Data Representation | Passive or Active | Quadrants | Year |
| --- | --- | --- | --- | --- |
| Data in context ... [44] | Spatial | Passive | | 2020 |
| Exploring spatial meaning ... [88] | Abstract | Passive | | 2017 |
| Augmented urban models ... [7] | Abstract | Passive | | 2025 |
| Augmented scale models ... [68] | Abstract | Passive | | 2022 |
| Urp: a luminous ... [86] | Spatial | Active | | 1999 |
| Illuminating clay ... [61] | Spatial | Active | | 2002 |
| Bringing clay ... [35] and | Spatial | Active | | 2004 |
| Real-time landscape model ... [55] and | Spatial | Active | | 2004 |
| Simultaneous worlds ... [31] | Spatial | Active | | 2022 |
| Manipulating tabletop ... [41] | Abstract | Active | | 2015 |
| CityScope ... [2] | Abstract, Spatial | Active | | 2018 |
| CityMatrix ... [94] | Abstract, Spatial | Active | | 2008 |
| Uplift: a tangible ... [23] | Abstract, Spatial | Passive, Active | | 2020 |

Table 1: Summary of situated visualisation work using tangible proxies and how they fit into the quadrants shown in Figure 4.

### 4.1 System Overview

Our system consists of three integrated components, enabling users to explore complex data and potentially collaborate in an intuitive, interactive environment:

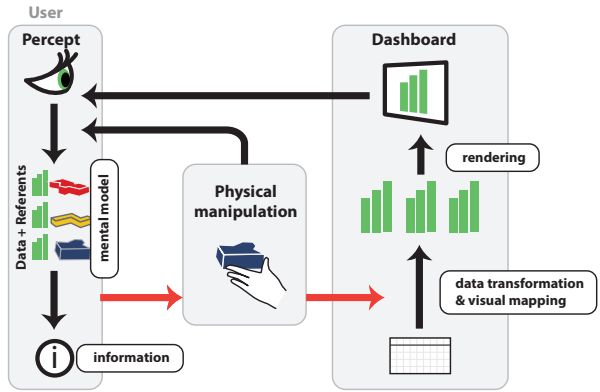

Figure 5: The interaction model of an active proxy interface in the context of situated visualisation. Red arrows represent user interaction. Black arrows represent percepts. The perception of physical manipulation and visualisation updates refine the mental model of binding between the data and referents, which is used to derive information. This diagram is inspired by Jansen and Dragicevic [38].

**Tabletop Display** display [23, 79, 87] provide large, horizontal interactive surfaces that support direct tangible interaction and collaborative data exploration. In our system, the tabletop display features a geographical map that serves as a platform for an autonomous tangible proxy. This central interface supports shared understanding through a collaborative model, incorporating natural interaction cues such as gaze tracking to indicate collaborators' attentional focus, deictic pointing gestures, and tangible object placement. Additionally, it allows direct interaction with the geographical map, including view angle adjustments and zoom level modifications.

**Autonomous Tangible Proxies** provide an engaging, intuitive, and interactive means of system interaction. The building model serves as a physical referent for embedded data visualisation, while physical interaction controls offer an intuitive method for collaborators to customise visualised data and layers. These autonomous tangible proxies are constructed by integrating tangible proxies with a custom-built desktop robot. By enabling autonomous turning and movement, the system eliminates the user effort required for repositioning proxies on the tabletop display after manipulation.

**Large Display Backdrop** provides space for abstract charts presented as a dashboard. This display is positioned adjacent to the tabletop display for convenience.

## 4.2 System Implementation

The 50-inch tabletop display presents a web application utilising Leaflet for visual interface implementation. A Microsoft Surface Hub serves as the large display backdrop. Autonomous tangible proxy tracking is conducted using a Vicon tracking system and a Unity application. The high-accuracy tracking system offers high flexibility for diverse design choices and broad scalability for future enhancements. However, it also introduces certain physical constraints that limit the interacting area and gestures. To minimise the impact of these constraints inherent to the tracking system, we used ten tracking cameras to maximise tracking coverage and angles. The desktop robot's surface incorporates five to six passive tracking markers to ensure accurate tracking and individual identification. The system components communicate through a local server built with standardised WebSockets, facilitating interconnection between different system instances. Figure 6 describes the setup.

*4.2.1 Autonomous tangible proxies.* Tangible proxies in our prototype are physical scale models of campus buildings. As proof of concept, we only printed five models of essential buildings. Five desktop robots with a diameter of 10 cm and weighing approximately 120 g are designed to carry the building models around on the tabletop display. Each robot is constructed with a compact configuration consisting of an Arduino Tiny ESP32-S2, a lithium polymer (LiPo) battery, a TB6612FNG dual motor driver, and two 50:1 micro gear motors, all housed within a 3D printed enclosure, as shown in Figure 7. A distinctive circular mounting area on the robot's upper surface is specifically designed to accommodate passive tracking markers, enabling precise localisation and identification through the Vicon motion tracking system. All robots are connected to the local server to receive real-time commands, which dynamically control the speed and direction of each wheel.

*4.2.2 Tangible proxies behaviour control.* This application controls the movement of the proxies and operates in conjunction with a Vicon tracking system, comprising eight Vicon Vantage V8 cameras and proprietary Tracker software. We created a 1:1 scale virtual environment in Unity for the tablet display and large display backdrop. We mapped the orientation and position data from Tracker onto this virtual environment. Each tangible proxy is a physical representation of a building and is linked to its respective geospatial position and orientation on the tabletop map display. The proxies are designed to be freely manipulated on the tabletop display and autonomously return to their corresponding position and orientation when not being interacted with. When a tracked proxy model is in touch with the tablet display, the Unity program checks whether it is at the correct map position and orientation. For proxies that are misaligned, we used Unity's built-in pathfinding algorithm to compute collision-free robot movement paths to guide them back to their corresponding destinations. The Unity application then transmits precise movement and turning commands to each robot.

Upon reaching its destination, each robot halts movement and rotates to match the required orientation. When a robot is repositioned on the tablet display or the interactive geospatial map changes, it recalculates a collision-free path to its corresponding map location. Safety boundaries are established around the tablet display's edges to prevent robots from inadvertently traversing beyond the interface. Robot pickup is detected based on the proximity between the proxy and the display. When a proxy is lifted, the Unity application immediately halts motor activity to ensure smooth physical manipulation. The system triggers different events based on the proxy's distance from the large display backdrop and its turning direction, sending corresponding commands to achieve various visualisation interactions described in subsection 4.4.

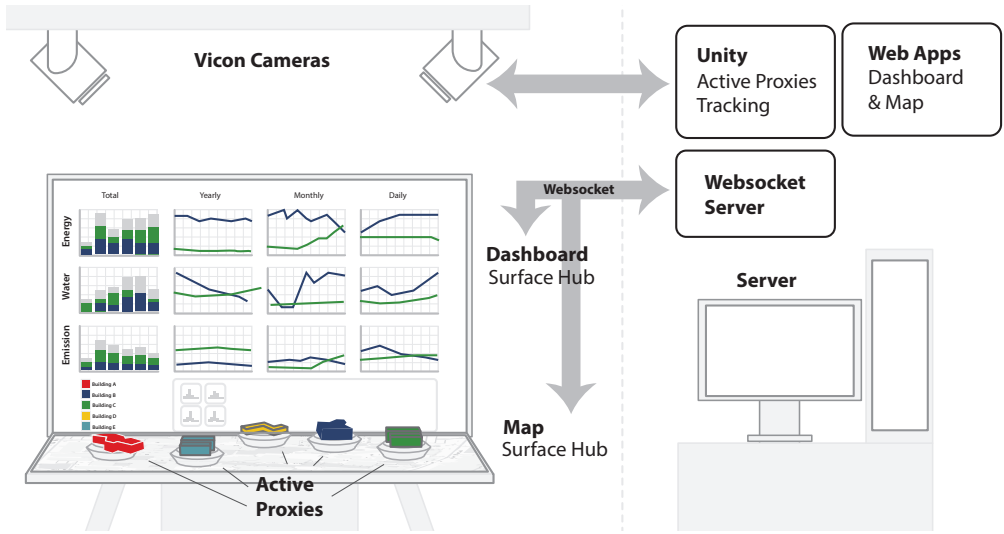

Figure 6: The technical setup of our MarioChart system consists of two Surface Hubs (for the dashboard and tabletop map), a Vicon System, and a server hosting the Unity application (for tracking and communications of the proxies), web applications (for the dashboard and map), and a WebSocket server.

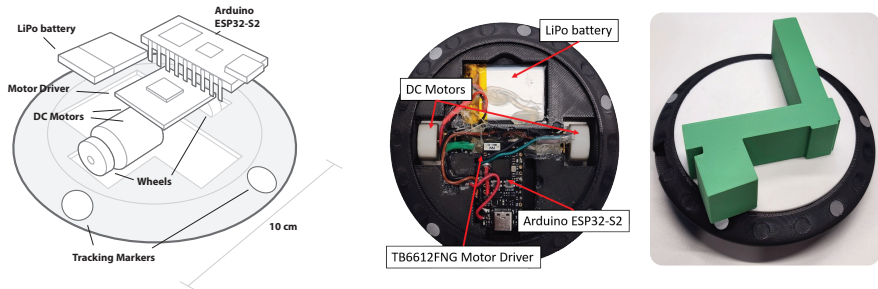

Figure 7: Left and middle: Components of our autonomous tangible proxy (the proxy is upside down). Right: A close look at the proxy (scale model and the self-driving base).

### 4.3 Workflow Design

There are many ways to pair physical manipulations and visualisation operators. At this stage, we designed a data exploration workflow that supports three key goals: 1) overview and detail, 2) filter and comparison, and 3) select and focus.

The *overview and detail* goal is to provide high-level information about the data across all buildings and the detailed information when needed. The default state of the dashboard—when all active proxies are placed on the tablet display—shows the overview data as the primary layer. In this state, all charts show aggregated values. To manage the multiple levels of data granularity, we implemented a secondary layer dashboard. This allows for a drill-down operation to explore progressively detailed information. For instance, the monthly electricity usage chart shows mean values across all buildings in the data, and drilling down from this chart revealed average electricity usage for each month.

The *filter and comparison* goal is the ability to inspect data of a specific building by filtering the charts. This is mainly achieved by pick-up and air-dwell physical manipulations. Picking up a proxy can be a clear indication of interest. Therefore, it makes sense to map this physical manipulation to the focus-related operators, such as filtering the entire charts on the dashboard. When a proxy is picked up, the charts are filtered by this referent. We highlight the general information of the referents on the building information panel. This filtering interaction supports multiple referents, allowing for data comparison between referents. Placing the active proxy removes it from the filter list, which eventually brings back the overview when the list is empty.

The *select and focus* goal is to reduce the users' information load by allowing them to focus on charts of interest. The selection is performed by picking up a proxy and then moving it near the intended chart. After a short dwell time, the chart is then selected and displays the detailed information as a secondary layer. Moving the proxy away from the chart moves it to the default primary layer. Users are also allowed to focus on selected charts for a specific active proxy. Selecting the chart requires users to pick up a proxy and flip it towards the target chart. To focus on previously selected charts, the user needs to pick up and flip the active proxies towards the shoebox panel (a shoebox is an abstract information storage, a term used in sensemaking research [3, 19, 62]). This select and focus interaction can also be combined with a filter and comparison. See the supplementary video for further demonstration of all interactions.

## 4.4 Interaction Design

*4.4.1 Fundamental Actions.* Our prototype initially supported the following applications, which we consider fundamental to initiating abstract data exploration. We mapped the physical actions performance with the tangible proxies to the manipulations of the visualisation chart, guided by Brehmer and Munzner's *Multi-Level Typology of Abstract Visualisation Tasks* [12]. The same taxonomy was also used to rank our designed experiment questions, allowing us to assign difficulty levels to each question type, see more in subsection 5.2.

**Picking Up/Placing Down**: Physical manipulation through picking up or placing down enables building selection for data inspection (following [23]). When one or multiple proxies are picked up, they are emphasised in the legend and visualisation charts, while non-selected proxies are dimmed, which *Selects* buildings and *Filters* the chart (Figure 8). Returning a proxy to the tablet display removes its highlighted effects.

**Drilling Down in Time**: Positioning proxies near the large display backdrop within a predefined distance threshold allows users to inspect specific category-time charts in greater detail, see Figure 9-a. For example, inspecting the chart of average yearly electricity consumption shows the average electricity consumption chart for each year. Maintaining the proxy within this distance threshold preserves the detailed view, and taking the proxy out of the distance threshold turns the visualisations back to the default view. This operation is equivalent to *changing* the visualisation chart to a different time-scale view based on our designed setting.

**Shoebox**: Pitching the proxy towards a chart enables users to *select* that chart for bookmarking using the shoebox (Figure 9-d). Within the shoebox panel, added charts are grouped by building, allowing users to quickly discern the total number of charts included for each building (Figure 9-e).

*4.4.2 Pilot study feedback and interaction elicitation.* We invited eight participants, including six visualisation researchers and two facilities managers (data domain) experts, to evaluate our autonomous active proxy system and provide feedback. Despite general enthusiasm about the system design, we received critical insights regarding the visualisation web interface. Participants consistently reported confusion due to the display's extensive visual complexity. With 12 visualisation charts spanning the display area, many participants expressed difficulty in determining focal points. Specifically, when attempting to drill down into the temporal details of a specific chart, users found it unclear whether their selection was successful. Additionally, maintaining a detailed view required users to hold the proxy within a distance threshold, which induced significant physical fatigue.

*4.4.3 Elicited Interactions.* Considering the feedback from our pilot study, we subsequently decided to introduce the following enhancements, with a particular focus on improving visual cues to elevate the overall interaction experience.

**Proxy Shadow**: Participants in the pilot study reported that the lack of visual cues when interacting with the tangible proxies makes it difficult for them to focus on the visualisation display, which contains multi-dimensional information. Several participants noted uncertainty about where to direct their attention, as one reported "I do not know where I should look at". Although the "Drilling Down in Time" interaction provided a visual highlight once the tangible proxies are held within the distance threshold, there were no visual cues before reaching that threshold. The absence of early feedback led some participants to question whether the system was operating correctly. To facilitate navigation, reduce learning complexity, and provide immediate feedback during interactions, we designed a shadow of the held tangible proxy to represent the proxy's projected position on the visualisation display. When a proxy is picked up, the corresponding building shape appears for easy identification, and its on-screen position follows the proxy's movement, providing a visual cue for navigation on the display. We further enhance the visual hint of the relationship between the proximity of the held tangible proxies and the visualisation display by using a magnifier icon which dynamically resizes based on the distance of the proxy to the display, intuitively guiding users to move the proxy closer to the display for details.

**Selection Highlight**: Users can select visualisation charts by moving a proxy in proximity to their desired chart. The proxy shadow provides an intuitive interaction, allowing users to visually confirm their selection through the direct overlapping of the proxy shadow with the target chart. To enhance user experience, we implemented a highlight animation that not only emphasises the selected chart but also serves as a visual cue. Specifically, we designed a flip animation that highlights the selection and implicitly suggests to users that they can flip the proxy in the same direction to add the chart to the shoebox.

**Lock/Unlock**: Following time-based drilling, users can lock a detailed visualisation by rotating the proxy clockwise (Figure 9-b). During this locked state, proxy interactions are limited to highlight modifications. Users can unlock the display at any time by rotating any proxy counterclockwise (Figure 9-c). The proxy shadow's icon dynamically represents the current lock state, with a directional arrow indicating the lock and unlock actions.

**Visual Cues for Shoebox**: To provide clear visual feedback for shoebox interactions, we incorporated several cues that indicate which visualisation has been added and its associated building. When a chart is added to the shoebox, a small building-shaped icon appears in the top-right corner of the corresponding visualisation. Simultaneously, an animated rectangle moves from the chart's original position to the shoebox panel, visually representing the feature of adding a chart to the shoebox.

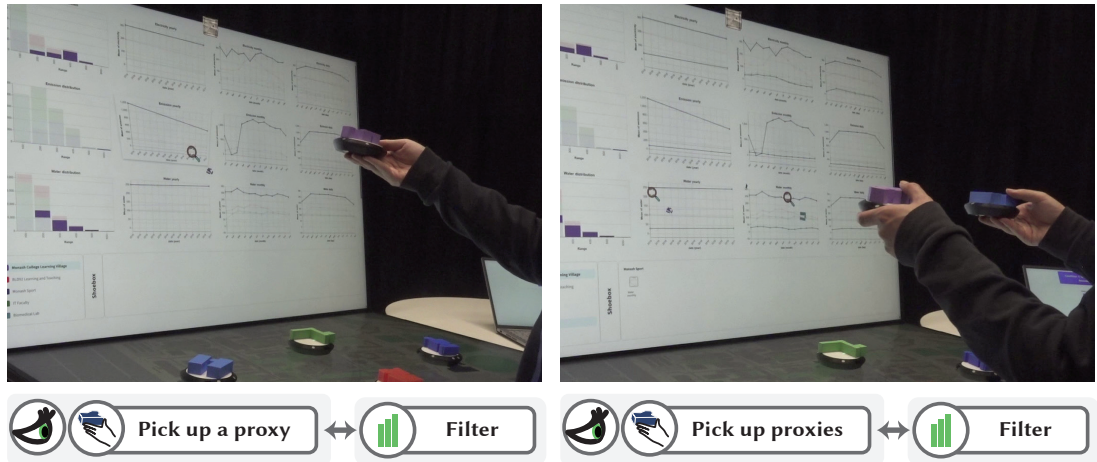

Figure 8: Picking up a proxy or multiple proxies is the basic physical action that our design employs to filter the charts on the dashboard.

### 4.5 Visualisation Design

The data representations are displayed on a Microsoft Surface Hub. Details are as follows:

*Data Representation* The charts are an essential component of the autonomous active proxies approach. In our implementation, we show 12 charts derived from three temporal attributes: electricity consumption, emission, and water consumption. Each row represents a single attribute presented at different levels of detail (e.g., a histogram for value distribution and line charts of yearly trend, monthly trend, and weekly trend).

*Legend* A comprehensive legend is included to display each building's name alongside its corresponding colour. When a building is highlighted in the visualisation, its corresponding entry in the legend is simultaneously highlighted.

*Shoebox* The shoebox panel is placed at the bottom of the dashboard. Charts are presented as small icons arranged in a grid layout.

We support the shoebox approach by allowing the user to select certain charts for each referent. Each referent can have a different set of charts. For example, energy consumption is relevant for the Teaching and Learning building with extensive air conditioning systems, while water consumption is more relevant to the Sports Building with swimming pools. The information stored in the shoebox panel is visualised in the dashboard. We also made this information persistent so that it can be reused throughout multiple analysis sessions.

## 5 User Study

We conducted a controlled user study to evaluate the active proxy concept through the MarioChart system. The study received ethical approval from the local ethics committee.

### 5.1 Dataset

The visualisation environment was designed for urban data exploration. It included five tangible proxies representing selected buildings on a map and a dashboard displaying a series of visualisation plots. The dataset was obtained from the University's Building and Management Department and contained information on energy consumption and generation. Because the available data did not fully cover the categories of interest, we restructured it into three categories: 1) electricity consumption, 2) emissions, and 3) water consumption, spanning the years 2016 to 2017. This ensured that the user study was based on realistic data variations.

### 5.2 Study Factors

Our main factor is the **Interface Type**. **Active Proxy Interface** is the MarioChart system we have described above. As a baseline, we develop a graphical user interface on a tablet. We refer to this level as **Tablet Interface**. Following the same interaction design mentioned in subsection 4.4, the user selects and filters building/chart with button-based interactions. A long press on the touchscreen enables the drill-down feature, while a dedicated button allows users to add a chart to the shoebox based on the current building selection. The dashboard can be locked after drilling down to the secondary layer, and users can unlock it with a single button press.

The other factor is **Task Types**. In our study, participants were asked to interact with the situated visualisation environment to answer 12 predefined data-analytic questions. These questions were organised into four categories, each reflecting a different level of difficulty based on Brehmer and Munzner's *Multi-Level Typology of Abstract Visualisation Tasks* [12] (see Table 2) and type of analytic task:

- **BM (Bookmark Questions)**: Participants were asked to bookmark three specific charts on the dashboard for a given

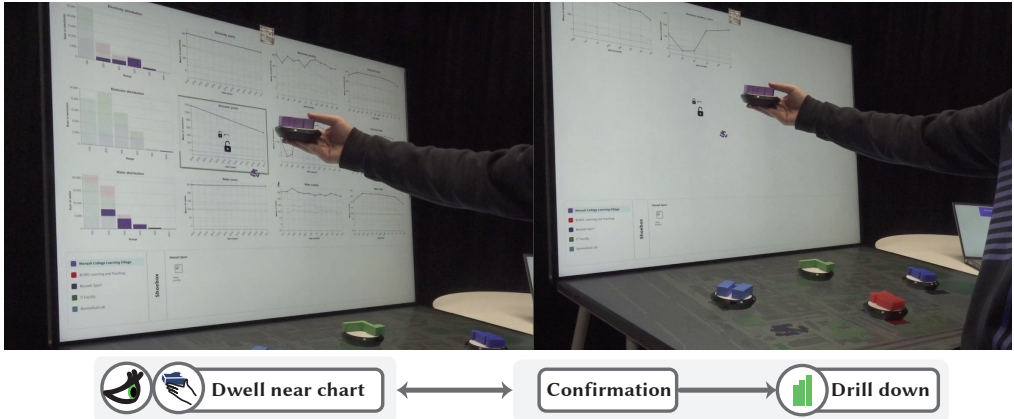

(a) Picking up and dwelling (left) a tangible proxy in the proximity of the intended chart for a short dwelling time to show more details (right)

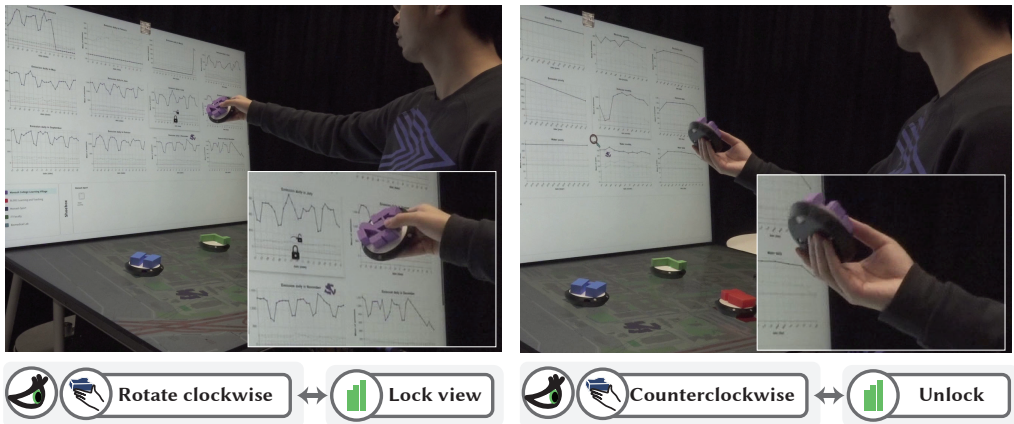

b) Rotating the proxy clockwise locks the dashboard from returning back to default charts while moving the tangible proxy away.

c) Rotating the proxy anti-clockwise unlocks the dashboard.

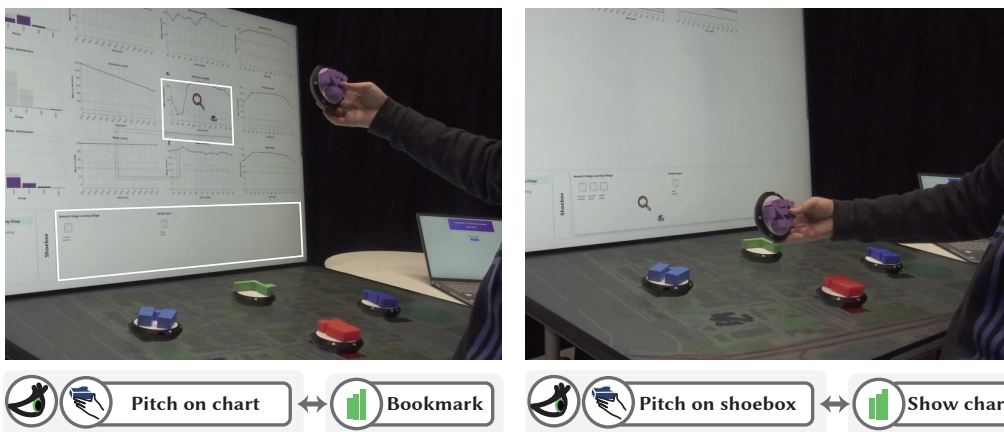

d) Pitching the proxy while pointing to chart bookmarks the chart to the shoebox.

e) Pitching the proxy while pointing to the shoebox shows all charts in the shoebox associated with the building.

Figure 9: More interactions with charts in the MarioChart system, apart from filtering in Figure 8.

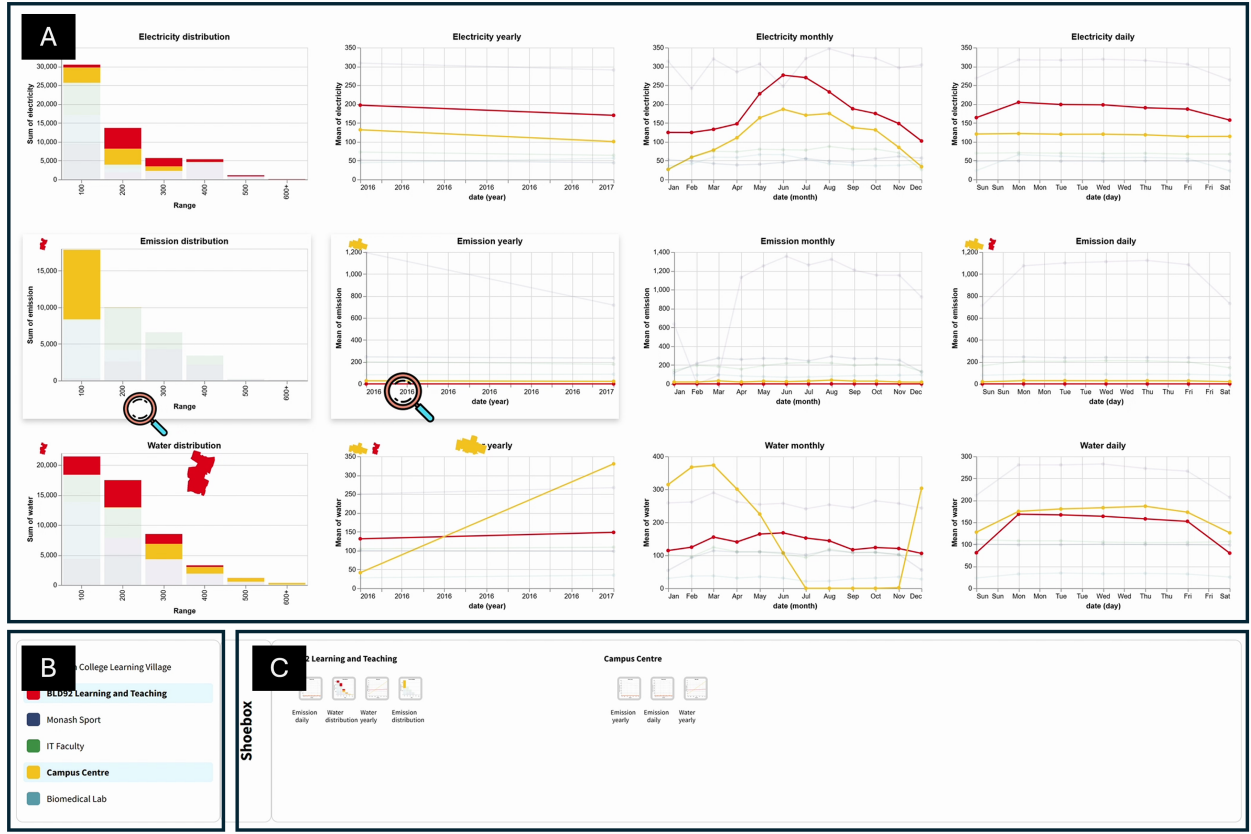

Figure 10: The dashboard of our implementation of Autonomous Active Proxies consists of the charts panel (A), legend (B), and the shoebox (C). In this screenshot, two proxies are picked up, triggering the charts' filtering.

building. This category was designed to assess basic navigation skills in locating required visualisations.

- **DR (Data–Referent Questions)**: These questions provided known data but an unknown referent, testing the participant's ability to identify the correct referent based on the given data.
- **RD (Referent–Data Questions)**: These questions provided a known referent but unknown data, requiring participants to locate the correct data associated with the referent.
- **DR-S (Data–Referent with Spatial Requirement Questions)**: These questions combined known data with a spatial requirement but with an unknown referent. This category tested participants' ability to identify the correct referent using both data and the building's spatial information on the map.

The four categories were designed to cover a broad range of analytic actions, ensuring participants experienced diverse interaction scenarios in both interfaces. Importantly, participants were not informed of the category labels (BM, DR, RD, DR-S) during the study. Although the specific questions differed across the two interfaces, the distribution of question types remained consistent, with three questions per category. Each interface focused on three buildings, with one building overlapping across interfaces. The other two buildings in each interface were unique and later used for comparing recall performance between interfaces.

| Question Type | Actions Required | Diff. Level |
|---|---|---|
| BM | Select Building → Filter by Building → Select Chart | 3 |
| DR | Select Chart → Change to Secondary Layer → Filter by Building | 3 |
| RD | Filter by Building → Select Chart → Change to Secondary Layer | 3 |
| DR-S | Select Chart → Change to Secondary Layer → Filter by Building → Navigate Building on the Map | 4 |

Table 2: Question types, action required and their difficulty levels.

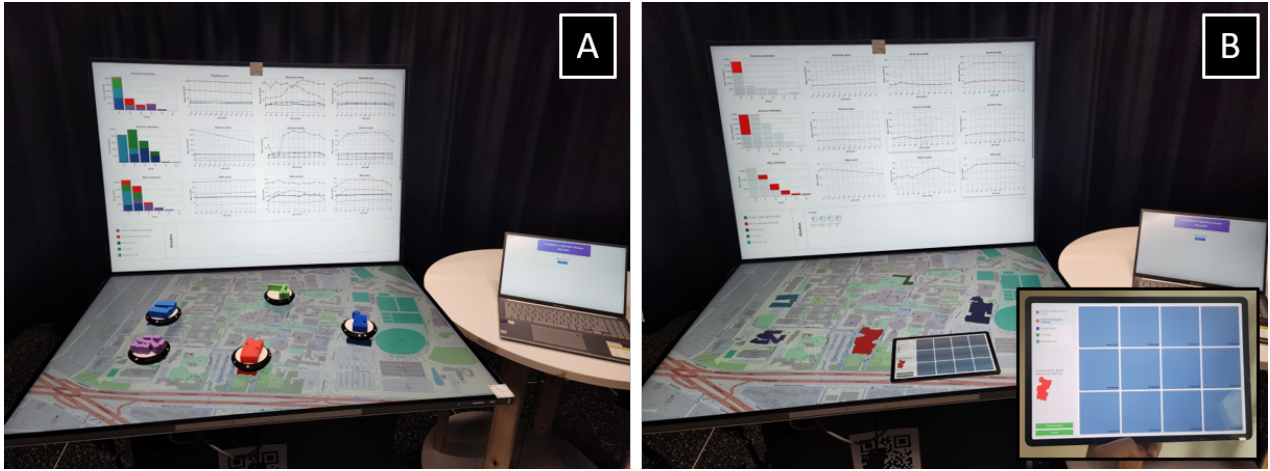

Figure 11: User study setup: (A) active proxy interface; (B) tablet interface.

To summarise, our study is a multifactorial 2 (Interface: Active Proxy, Tablet) × 4 (Task: BM, DR, RD, DR-S) within-subject experiment. The order of the Interface was counterbalanced across participants to mitigate potential learning effects, and the order of task type was randomised across interfaces to reduce the bias from predictable sequences. In total, 12 participants were recruited from our university (8 male, 4 female; mean age = 24.3, median age = 22).

### 5.3 Procedure and Data Collection

**Phase 1** of our study was run in our lab, where participants performed the tasks with the two interfaces. At the start of the study, participants received a brief introduction to the setup and background of the situated visualisation environment. We collected consent and demographic data. After the introduction, we started the experiment. For each interface, participants were given time to practise interacting with dummy data until they felt comfortable proceeding to the formal study tasks.

For every question, the system recorded both the response timestamp (in milliseconds) and the submitted answer, which was later evaluated for correctness. During the experiment, participants answered the 12 questions sequentially, viewing only one question at a time. After submitting an answer, the system automatically went to the next question. Their interactions were video filmed and observed by the experimenter to identify potential behavioural patterns. In the Active Proxy interface, physical proxies representing the selected building models were placed on the map for participants to interact with, see A) in Figure 11. In the Tablet Interface, the proxies were removed, and participants instead used a Samsung tablet with a custom-developed touchscreen interface for data manipulation, see B) in Figure 11. The dashboard comprised 12 visualisations in total. For each category, it presented one distribution chart and three average-value charts at different temporal scales: yearly, monthly, and weekly.

At the end of each interface, participants completed a set of questionnaires, including one Borg CR10 fatigue item [11], one PAAS mental load item [59], 12 items from the User Engagement Scale – Short Form (UES-SF) [58], three open-ended immediate recall questions, and two qualitative feedback items (one on perceived advantages and one on perceived drawbacks). After completing both interfaces, participants were asked to indicate their overall preference and provide additional qualitative feedback. They were then invited to schedule an online interview with the experimenter at least 24 hours after the study to capture their responses for long-term recall. The first phase of our study takes approximately 50 minutes to 1 hour.

**Phase 2** of our study looks at long-term information recall. Similar to Bateman et al.'s *Useful Junk* study [6], we asked questions during the interview and manually recorded participants' long-term recall responses. However, unlike Bateman et al. study, participants in our study were not asked to read the visualisations in detail, as multi-variate visualisations often contain too much information. Instead, the goal of our study was to answer analytic questions related to urban data visualisations. Therefore, our recall test focused primarily on the geo-position of buildings and obvious patterns visible in the charts, such as peaks or trends. To capture this, we used open-ended questions for both immediate and long-term recall, and scored responses based on the number of points correctly matched to the data (details of the scoring in subsection 6.3). During the interview, the experimenter asked participants to recall a building's label based on a grey-coloured building shape. This was followed by questions about the building's relative position and any associated data, such as peaks, trends, or other notable patterns. If a participant's response was too vague, for example, missing a description of the data category or time range, the experimenter prompted them with clarifying questions such as: "Can you be more specific on X?", where X referred to the missing information (e.g., "the category of the peak/trend" or "the time range"). Phase 2 took approximately 10 minutes.

## 6 Result

We first present the performance results for each interface, followed by perceived load (physical and mental), user engagement, recall scores, and finally qualitative feedback. For analysis, we used bootstrap confidence intervals and t-tests in most cases, while Wilson scores were applied to binary correctness data. In line with current best practices, we pre-registered our empirical analysis [9, 18] and made all experimental materials, data, pre-registration, and code openly available at [redacted for anonymity]. Below, we detail the results of this pre-registered analysis.

## 6.1 Performance

We report the absolute mean completion times (in seconds) across interfaces and question types (see Figure 12a). To address the positive skewness of completion time data, we analysed log-transformed values and present anti-logarithmic results. Given the log-normal distribution of time data, this approach is a standard practice in HCI and visualisation research [8, 39, 89]. As shown in Figure 12a, we could not find evidence that would indicate that Active Proxy or Tablet Interface has better performance in overall completion time. When comparing across question types, we found strong evidence that for questions of the RD type, Active Proxy (72.40, CI [60.54, 86.58]) outperformed Tablet Interface (95.56, CI [81.64, 111.86]), Cohen's d = -0.538. The effect appears to be quite strong, allowing participants to save 23.16 seconds. We report the mean correctness (in %) for each interface and question type (see Figure 12b). We could not find evidence of a difference between Active Proxy and Tablet Interface in terms of correctly answering the data analytic questions. We also analysed the mean completion time for correctly answered questions and found similar pattern to Figure 12a.

## 6.2 Perceived Load and Engagement Results

We report the estimated physical load (see left in Figure 12c), measured using the Borg CR10 scale of physical fatigue, and the estimated mental load (see right in Figure 12c), measured using the PAAS scale of cognitive load, for each interface. We could not find evidence of a difference between Active Proxy and Tablet Interface in terms of either physical or mental load. We also analysed the mean user engagement scores, both overall and by subclass, for each interface. Similarly, no statistically significant differences were observed between Active Proxy and Tablet Interface in user engagement scores.

## 6.3 Recall

To evaluate participants' recall responses, we first create a scoring checklist that lists all visible data patterns, such as peaks or trends, as bullet points for each building, time range, and data category. Each recall response is then broken down into points corresponding to each building and compared against the checklist. Recall scores for data-related information and geo-position information are stored separately for later analysis. Scoring is assigned as follows:

- **1 point (correct):** The participant's recall response matches a point in the checklist.
- **0.5 point (partially correct):** The participant's recall response partially matches a point in the checklist but is missing some information, such as the category or time range.
- **0 point (all incorrect):** The participant's recall response does not match any point in the checklist.

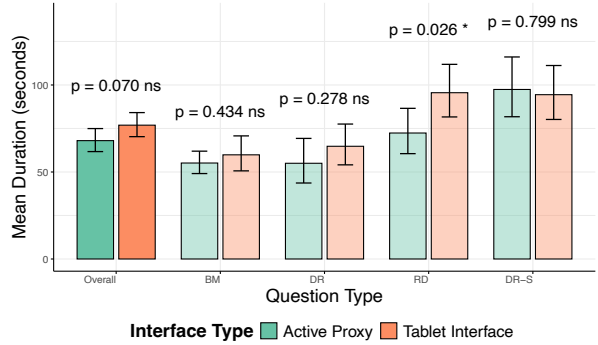

(a) Completion Time

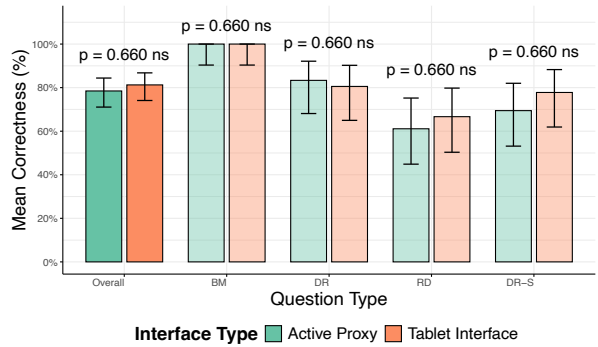

(b) Correctness

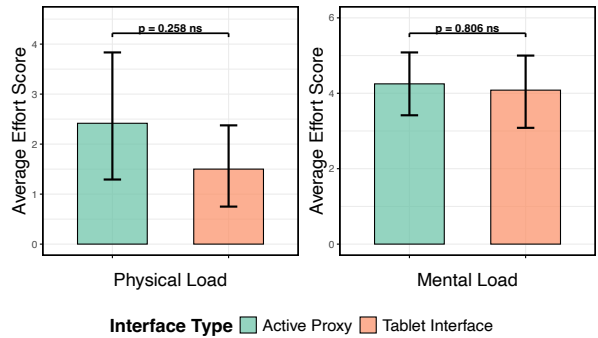

(c) Left: Mean score from Borg CR10. Right: Mean score from PAAS. Error bars are 95% Confidence Intervals (CIs).

Figure 12: Performance data for each interface and each question type. Error bars are 95% Confidence Intervals (CIs).

- **0 point (I can't remember):** The participant is unable to recall any information.

Using the recall scores, we report the average recall score (see a) in Figure 14) as well as the two subset scores, data recall (see b) in Figure 14) and position recall (see c) in Figure 14), for each interface and for different recall types. There is strong evidence that the Active Proxy (1.173, CIs[0.736, 1.638]) has higher immediate recall scores than Tablet Interface (0.493, CIs[0.250, 0.798]), Coden's d = 0.706. Specifically, immediate data recall does not appear to exhibit differences based on our data, whereas we found strong evidence that immediate position recall scores are higher for Active Proxy (0.333, CIs[0.167, 0.500]) comparing to Tablet Interface (0.083, CIs[0.00, 0.188]), Coden's d = 0.746. We did not find evidence of differences for long-term recall. We examined the correctness of participants' recall of building names based on the corresponding grey-coloured building shapes ( see d) in Figure 14). No evidence indicates that either Active Proxy or Tablet Interface improves immediate or long-term memory for mapping building names to building shapes.

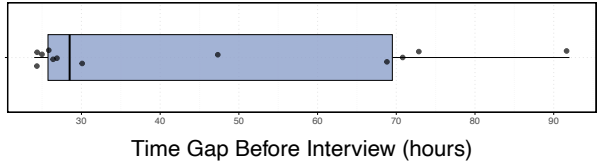

**Figure 13: Distribution of time gaps before long-term recall interview across participants.**

### 6.4 Preference and Qualitative Feedback

We present the preference and the qualitative feedback gathered from participants regarding the different interaction interfaces. Among the 12 participants, seven expressed a preference for Active Proxy, while five preferred the Tablet Interface.

We also report the perceived advantages and disadvantages of each interface based on participants' comments. For Active Proxy, nine participants described it as "interesting" or "fun", highlighting the autonomous proxy's ability to self-relocate on the map. Several participants appreciated this feature with comments like *"I like the way that the proxies returned back somewhat to their rightful place in the map."* or *"the buildings returning to their locations is very cool."*. Five participants noted that interacting with tangible proxies for data manipulation felt "intuitive". However, six participants commented that holding the tangible proxy for extended periods required considerable "physical effort" and was "tiring" with remarks such as *"It's hard to hold for this proxy manipulation."* or *"Holding the proxies is tiring."*. Additionally, four participants mentioned that they sometimes "unintentionally" triggered functions during interaction. For the Tablet Interface, nine participants used phrases such as "easy to use" and left comments like *"It was more familiar to me."* and *"It is easy to control."*, and six emphasised that it allowed them to perform accurate actions with ease, as one noted, *"It's easy to use to find more details through [Interactions]."*. Reported drawbacks were minimal: aside from general user interface improvement suggestions (not directly relevant to our study goals), only one participant noted the inconvenience of frequent focus switching between the tablet and the dashboard screen. Two participants explicitly stated after the study that they could not identify any negative aspects of the touchscreen interface.

### 6.5 Interaction Analysis

We also analysed participants' interaction behaviours across the two interfaces using the user-study video recordings. For each participant, we manually annotated the video segments and categorised them into three types of behaviour: (1) the interactions with the interface, (2) the periods focused solely on reading the data, and (3) the instances where participants needed to double-check the task description. Those behaviours were further decomposed into the detailed actions described in Table 3. In line with prior work that examines interaction behaviour through number of operations [33], we used the count of repeated actions to reflect the frequency of interaction patterns, and the duration of each action to assess the efficiency of participants' interactions across both interfaces. Quick interactions (less than 3 seconds), such as filtering and locking, were annotated as instant actions, whereas slower interactions were annotated only when participants took at least three seconds performing them. This threshold allowed us to exclude brief, unintended movements and focus on meaningful interaction behaviours. We aggregated the count of actions and their duration to perform statistical and visual analysis.

| Code | Reflected Behaviour |
|---|---|
| Question Confirmation | Participant rereads the task description to verify or clarify the required action |
| Filter | Participant applies a filter to the visualisation to show data for specific buildings |
| Unfilter | Participant removes a previously applied building filter |
| Chart Selection | Participant selects a chart and drills into its secondary layer to view more detailed information |
| Reading Data | Participant spends time examining and interpreting data shown on the visualisation display |
| Lock | Participant activates the lock function to fix the visualisation display in its current state |
| Unlock | Participant deactivates the lock function to resume interactions with the visualisation |

**Table 3: Annotation codes used in the interaction data analysis.**

Apparent patterns shown Figure 15 a) are the filter and unfilter actions. We observed a slightly, although not statistically significant, lower filtering actions performed with the Tablet Interface (4.51, CIs[3.71, 5.41]) compared to the Active Proxy interface (5.58, CIs[4.36, 6.96]), Cohen's d = 0.275. For unfiltering actions, the Tablet Interface (3.81, CIs[3.14, 4.49]) showed a significantly lower mean count than the Active Proxy interface (5.51, CIs[4.21, 6.98]), Cohen's d = 0.452. All other annotated action counts showed no notable differences across interfaces. Figure 15 b) indicates that participants spent significantly longer reading data in the Tablet Interface (13.76, CIs[11.45, 16.57]) than that in the Active Proxy interface (10.12, CIs[9.02, 11.28]), Cohen's d = -0.565, while the time spent confirming tasks and completing chart selections was similar for both interfaces. Additional plots of the mean counts and durations for each question type are provided in the supplementary material.

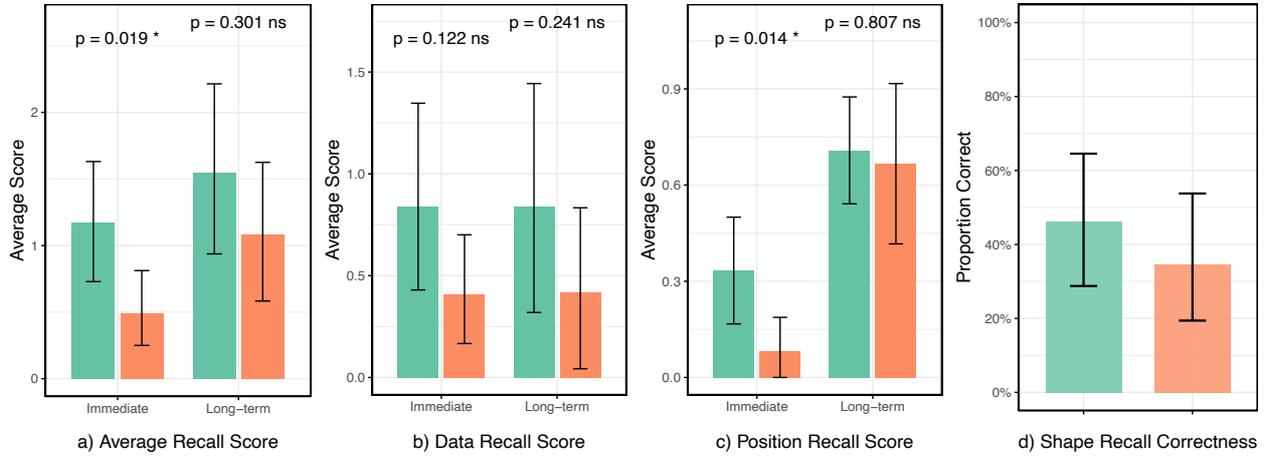

Figure 14: a) Average Recall Score including both data recall and position recall; b) Data Recall Score; c) Position Recall Score; d) Correctness of Shape Recall.

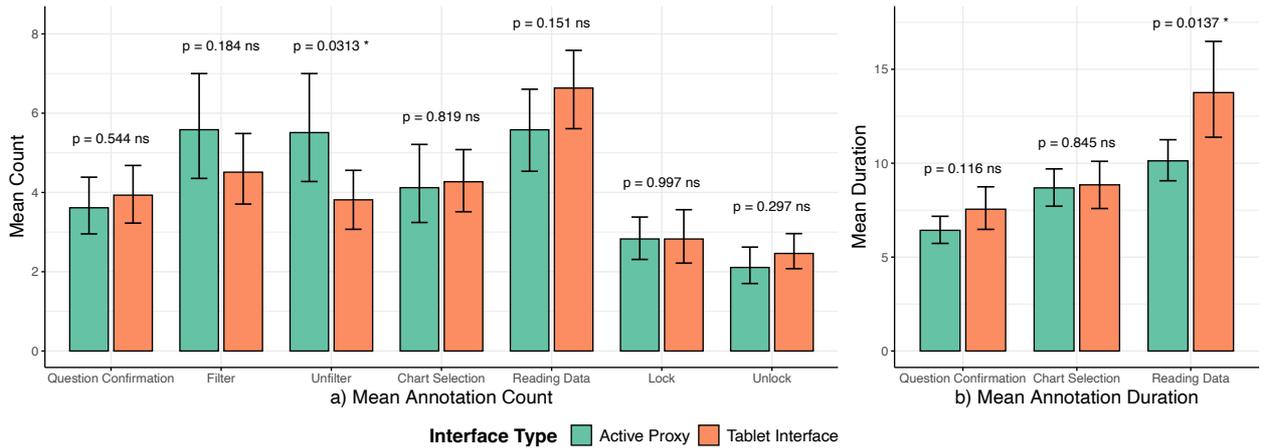

Figure 15: Mean count and duration for each interface and each annotation type. Error bars are 95% Confidence Intervals (CIs).

## 7 Discussion

### 7.1 Information Recall

Motivated by research in TUIs, we hypothesised that Active Proxy could strengthen the user's information encoding between the referent and abstract data representation. Our study did not find any significant evidence supporting this. There is no difference in data recall between Active Proxy and tablet interfaces in immediate and long-term recall. Similarly, the shape recall – which does not include any information related to the data – did not show any significant differences for both immediate and long-term recall. We found that participants primarily relied on the colour to identify data referents, paying little attention to shape. This is evident by our observation in phase 1 and the fact that most participants recall the colour first when prompted with the building shape in phase 2. This may suggest that attention paid to the building shapes and labels is overshadowed by the reliance on the colour encoding of the referent.

We found significant differences in position immediate recall, where the Active Proxy interface is superior to the tablet interface. Though this is not the case for long-term recall. The spatial-information recall accuracy of the Active Proxy interface aligns with existing work in TUIs where tangible interactions were found to benefit the user's spatial cognition [43, 48, 60, 75]. This suggests that the benefit of tangible interface and interaction in spatial cognition is likely to extend to a situated data visualisation context with physical manipulation of tangibles.

## 7.2 Performance

Our study results suggest that overall, the Active Proxy interface is comparable with the graphical user interface on a tablet but is significantly superior in a certain task. We only found a significant difference in performance on the Referent-Data question. This question requires users to locate the correct data associated with the referent from a known referent. This observed time advantage could be attributed to the more direct access to the referent in the active proxy interface, as referents are configured consistently on the map according to their relative positions. Accessing referents on the graphical interface on a tablet, on the other hand, relies on visual scanning and filtering to identify targets. This interpretation is further supported by higher immediate position recall scores for the active proxy. Despite the limited performance benefit of the active proxy interface, this result aligns with existing work in TUIs demonstrating the performance advantages of TUIs over GUIs [24, 37, 80, 83, 93, 95].

## 7.3 Physical and Mental Load

In terms of physical load, we did not find any significant difference between the two interfaces, with both averaging between 1.5 and 2.5 on the Borg CR 10 scale, indicating a weak physical load. From the participants' feedback, we notice several interesting points. For Active Proxy, participants reported that fatigue was most pronounced when they needed to hold a tangible proxy above shoulder level to interact with visualisation charts, especially those located in the top corners of the dashboard. The required effort can be expected to increase with the height of the visualisation workspace, as task height is a primary determinant of physical fatigue [56, 77], and is further moderated by individual factors such as arm length and viewing distance [70]. Notably, every physical action in Active Proxy was inherently tied to the answer-finding process, as interactions with the tangible proxies occurred only when participants actively sought information. In other words, the physical effort, such as grasping, moving, or positioning a proxy, was always meaningful and directly advanced task completion. In contrast, some of the physical effort in the tablet interface did not contribute to solving the task. Similar to physical load, we did not find any significant difference in mental load, with both interfaces averaging around 4 at the PAAS score. Overall, findings on physical and mental load from our study expand the current literature, which currently does not explore these measures in depth. Our study suggests that despite the physical manipulation, the active proxy interface does not require high physical and mental demand.

## 7.4 Novelty and Interest

Qualitative feedback further suggests that an Active Proxy system, such as MarioChart, can attract user interest, provided that the interaction design is intuitive and seamlessly supports the analytic tasks. This observation aligns with prior research on tangible and embodied interaction, which emphasises that the usability and adoption of such systems depend less on their novelty and more on how naturally the interaction paradigm maps to users' expectations and task requirements [29, 36]. The positive feedback reported in subsection 6.4 highlights the potential of tangible systems to serve as viable alternatives in data visualisation contexts, particularly when intuitive mappings can transform physical effort into meaningful analytic engagement.

## 7.5 Interaction Pattern

Our analysis reveals a few interaction patterns distinguishing the Active Proxy interface and the table interface. We found a potential trend in which users engage more actively when using Active Proxy interface to perform filter and unfilter actions by picking up and putting down the proxies. This observation aligns with the trend of higher physical load observed in the Active Proxy interface (Figure 12c). Conversely, the shorter time spent reading data in the Active Proxy interface may indicate more efficient information comprehension, which is consistent with the better recall scores for Active Proxy shown in Figure 14. Although other behavioural patterns show similar patterns across interfaces, we believe additional differences may emerge with a more targeted study design. For example, our observations suggest that filtering by building in the Active Proxy interface may be slightly more challenging, as participants appeared to need additional time to associate each building's name with its corresponding shape at the beginning of the Active Proxy interface, even though a detailed legend was provided.

## 8 Limitations

Our study has several limitations that should be acknowledged. First, the use of a camera-based motion capture system with Vicon for tracking the tangible proxies introduced constraints on how participants could hold the proxies. While the system provides high accuracy and offers strong potential for future applications that require precise tracking, its reliance on the cameras with an unblocked view of the reflective markers attached to the top of the proxies limits the holding postures. To mitigate this issue, participants were explicitly trained to hold the proxies in ways that minimised occlusion of the markers, which, although accepted by most, did not always correspond to their preferred or most natural grip. Alternative tracking methods, such as IMU-based tracking, could mitigate this limitation further by removing the need for such constraints. Nevertheless, the potential influence of required holding postures on user experience should be taken into account when interpreting the findings.

Second, the number of tangible proxies used in the study was limited to five due to hardware and time constraints. While this setup was sufficient to demonstrate the feasibility of our system and to yield meaningful insights, it does not fully represent the potential of a *proxsituated* visualisation environment. Ideally, such an environment would allow all referents to be represented as autonomous tangible proxies. Although our results suggest that Active Proxy can enhance users' understanding and impression of spatial data, it remains an open question whether these benefits would scale to larger numbers of proxies or whether interaction complexity and physical effort might impact the advantages observed.

Third, while a small sample size should be acknowledged as a limitation of a study, it is important to highlight that there is no magic number for participants required in a study [4], and visualisation studies often have small numbers of participants but still provide relevant results [15].

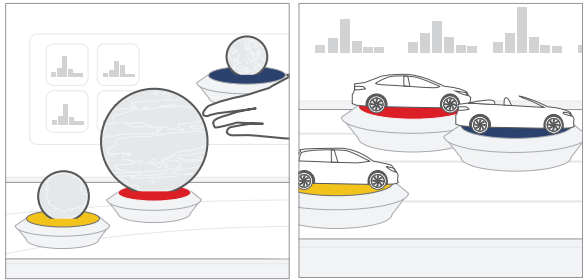

Figure 16: Potential use cases of the MarioChart system include solar system data visualisation and simulation (left) and cars dataset visualisation (right).

## 9 Additional Use Cases

Our current prototype is built based on the use scenario for campus building data exploration and analytics. We have also considered other potential use cases for our system can be used for.

**Solar System**. We propose an alternative use scenario that leverages our system to simulate celestial body movements[7], as shown on the left side of Figure 16. By defining the sun's position on the tablet display, we can model planetary movements, reflecting their relative spatial relationships. This visualisation allows for temporal exploration, enabling users to travel forward and backward in time to observe planetary positioning changes. The system can simulate space exploration activities by mapping the robots accordingly. For example, by designating separate proxies for Earth, Moon, and a space exploration rocket, a rocket trajectory in space can be simulated on the tablet display. The accompanying visualisation web interface supports data-driven comparisons, allowing exploration of planetary analytics such as size, mass, and rotation speeds.

**Cars Dataset Visualisation**. Our system could also demonstrate intuitive comparative visualisation by replacing building models with car models and their corresponding data, as shown on the right side of Figure 16. The web interface enables users to select and compare specific car models by manipulating the car model proxy. Users can compare selected models and focus on particular attributes of interest for each car model. The tablet display can be transformed into a racetrack where autonomous active proxies represent different car models moving in the racetrack according to each model's specific configurations, such as acceleration, maximum speed, and turning capabilities. This approach provides a spatial and interactive representation of car performance characteristics.

Other scenarios, such as consumers manipulating physical products or surgical trainers handling organ models to explore related information, are also good fit to this active proxy system.

## 10 Conclusion and Future Work

In this paper, we introduce and evaluate the concept of active proxies for situated visualisation, positioning them within a conceptual space defined by two axes: passive versus active proxies and spatial versus abstract representations. Our design and implementation of an active proxy interface, termed the MarioChart system, provides an opportunity to test this concept in practice and demonstrates its capability. Through an empirical study with 12 participants, we conducted an evaluation comparing the active proxy interface against a graphical user interface on a tablet. Despite the intertwined interaction between the user, referent, and data in the MarioChart system, we did not find significant evidence supporting the benefit of active proxy techniques in strengthening the binding between referent and data representation in users' minds. However, our findings suggest that active proxy interfaces can enhance short-term memory of spatial information regarding the referent and outperform tablet interfaces in analytical tasks requiring rapid access to referents. We also found no significant differences between active proxy interfaces and touchscreen tablets in long-term memory, physical and mental workload, or user engagement.

Our work offers an initial baseline for further exploration of situated visualisation with active proxy interfaces, which constitutes part of the four quadrants we have proposed within our design space. Future research in this area should deepen the evaluation of specific analytical tasks in which active proxies might prove beneficial. Future active proxy design should ensure that users attend to the shape of the referent object. Subsequent empirical studies need to ensure that shapes assume an important role in task completion.

## Acknowledgments

The authors wish to thank their participants. We also thank Prof. Ariel Liebman for his early inspiration and support of the Uplift project, which helped motivate this work. The work was supported, in part by the Marcus and Amalia Wallenberg Foundation (grant MAW 2023.0130) and the Knut and Alice Wallenberg Foundation (grant KAW 2019.0024).

---
[7]https://www.kaggle.com/datasets/iamsouravbanerjee/planet-dataset